\documentclass{webofc}

\usepackage[varg]{txfonts}   
\usepackage{hyperref}
\usepackage{url}
\usepackage{amsmath}

\newcommand{\Ds}{$\mathrm{D}_{s}^{\pm}$~}
\newcommand{\Lc}{$\mathrm{\Lambda}_{c}^{\pm}$~}
\newcommand{\Dzero}{$\mathrm{{D}^{0}}$~}
\newcommand{\Lctopkpi}{$\mathrm{{\Lambda}_{c}^{\pm} \rightarrow p^{\pm}k^{\mp} \pi^{\pm} ~}$}
\newcommand{\Dstophipi}{$\mathrm{{D}_{s}^{\pm} \rightarrow \phi \pi ^{\pm} \rightarrow (k^{+}k^{-}) \pi^{\pm}}$~}

\newcommand{\pt}{$\mathrm{p}_\mathrm{T}$~}
\newcommand{\vtwo}{$v_2$~}
\newcommand{\vthree}{$v_3$~}
\newcommand{\vn}{$v_n$~}
\newcommand{\sqrtsNN}{$\sqrt{s_\mathrm{NN}}$}
\newcommand{\RAA}{$R_{AA}$~}

\hypersetup{colorlinks=true,citecolor=blue,urlcolor=blue,linkcolor=blue}
\begin{document}
\title{Investigating charm quark interactions and hadronization in PbPb collisions with \Ds and \Lc measurements}
%
%

\author{\firstname{Nihar Ranjan} \lastname{Saha}\inst{1}\fnsep\thanks{\email{nihar.ranjan.saha@cern.ch}} 
}

\institute{Indian Institute of Technology Madras, India}

\abstract{Charm quarks serve as a sensitive probe of the Quark-Gluon Plasma (QGP), providing direct insights into its formation, evolution, and properties. In this proceeding, we present new high-precision measurements from the CMS experiment in Pb-Pb collisions at \sqrtsNN{} = 5.02 TeV. The elliptic (\vtwo) and triangular (\vthree) flow of prompt \Ds mesons are measured with high precision, extending the kinematic coverage. A comparison of the \Ds flow to that of \Dzero mesons is performed to investigate the impact of strange quark hadronization on charm quark collectivity. We further explore charm quark hadronization by measuring the \Lc nuclear modification factor (\RAA) as a function of transverse momentum for different collision centralities, and the \Lc/\Dzero yield ratio in pp, pPb, and Pb-Pb collisions. These results provide crucial constraints on charm quark energy loss and hadronization mechanisms, offering critical tests for theoretical models and advancing our understanding of heavy quark dynamics in the QGP}
\maketitle
\section{Introduction}
\label{intro}
The study of the Quark-Gluon Plasma (QGP), a state of matter consisting of deconfined quarks and gluons, is a primary focus of ultra-relativistic heavy-ion collisions \cite{QGP1,QGP2,QGP3}. Heavy quarks, namely charm and bottom, serve as exceptional probes of the QGP's properties. Produced in initial hard-scattering events, they traverse the medium throughout its evolution, interacting with its constituents. Consequently, their final-state kinematic distributions carry vital information about the mechanisms of parton-medium interaction, such as collisional and radiative energy loss, and the degree of their participation in the collective expansion of the medium.

Two key observables, the nuclear modification factor (\RAA) and azimuthal anisotropy coefficients (\vn), are central to characterizing these interactions. The \RAA quantifies in-medium energy loss, while coefficients like elliptic (\vtwo) and triangular (\vthree) flow reveal the degree to which heavy quarks participate in the collective expansion of the QGP. The interpretation of these observables is deeply connected to the charm quark hadronization process. In the QGP, hadronization may proceed via coalescence with light quarks from the medium, in addition to vacuum fragmentation. This makes different charm hadron species sensitive probes: the \Ds meson to the strange quark environment and the \Lc baryon to the baryon density. Comparing their behavior to non-strange mesons like the \Dzero is therefore crucial for disentangling parton collectivity from hadronization dynamics.

In this proceeding, we present new results from the CMS experiment that provide deeper insights into charm quark dynamics in Pb-Pb collisions at \sqrtsNN{} = 5.02 TeV. We report high-precision measurements of the \vtwo and \vthree of prompt \Ds mesons over extended kinematic ranges. By comparing the flow of prompt \Ds to \Dzero mesons, we explore the impact of strange quark hadronization on charm quark collectivity. Furthermore, we investigate the charm quark hadronization mechanism by measuring the \Lc \RAA in different centrality classes and the \Lc/\Dzero yield ratio across pp, pPb, and Pb-Pb collisions. These results, when compared with theoretical models, provide critical validation of the underlying physics and significantly advance our understanding of heavy quark interaction with the QGP.

\section{Analysis detail}
\label{Ana_detail}
This analysis utilizes the high-statistics dataset of Pb-Pb collisions at \sqrtsNN{} = 5.02 TeV recorded by the CMS experiment at the LHC. The exceptional tracking capabilities of the CMS silicon tracker are essential for the precise reconstruction of charged-particle trajectories and the identification of secondary vertices displaced from the primary interaction point, a key feature of charm hadron decays.

The prompt \Lc is reconstructed via the decay channel \Lctopkpi with branching ratio $6.28 \pm 0.32 \%$, while the prompt \Ds is reconstructed via the decay channel \Dstophipi , having branching ratio $2.27 \pm 0.08 \%$. A multivariate analysis approach ~\cite{tmva}, employing Boosted Decision Trees, is used to optimize the selection criteria for decay candidates. This technique enhances the signal significance by effectively discriminating against the large combinatorial background. The contribution from non-prompt charm hadrons, originating from b-hadron decays, is statistically separated from the prompt signal based on the candidate's decay length information.

Following signal extraction, the flow coefficients of \Ds are measured using a well-established scalar product (SP) method~\cite{SP-STAR}, which evaluates anisotropic flow by correlating the azimuthal angle of the particle with the event plane. The flow harmonic $v_n$ is calculated as:

\begin{equation} \label{eq:qsp} {v_n}\{ {{\rm{SP}}} \} \equiv {\frac{\left\langle {{Q_n^{D_{s}^{\pm}}}Q_{nA}^{*}} \right\rangle } {\sqrt {{\frac{\left\langle {Q_{nA}Q_{nB}^{*}} \right\rangle \left\langle {Q_{nA}Q_{nC}^{*}} \right\rangle } {\left\langle {Q_{nB}Q_{nC}^{*}} \right\rangle }}} }}. \end{equation}

Here, $Q_{nA}$ and $Q_{nB}$ represent the flow vectors constructed from the hadron forward (HF) calorimeters in the pseudorapidity regions $-5 < \eta < -3$ ($HF^-$) and $3 < \eta < 5$ ($HF^+$), respectively. The $Q_{nC}$ vector is computed using tracks in the midrapidity region of the tracker ($|\eta| < 0.75$), while $Q_n$ denotes the flow vector associated with each \Ds meson candidate. Event-averaged correlations such as $\langle Q_n Q_{nA}^{*} \rangle$ are used to quantify the anisotropy, with the average taken over all \Ds candidates in each event class. To suppress non-flow effects and avoid autocorrelations, the subevent $A$ is chosen from $HF^-$ when the \Ds meson lies in the forward region ($\eta > 0$), and $HF^+$ is used when the \Ds lies in the backward region ($\eta < 0$), maintaining $|\Delta\eta|>3.0$. The flow vectors (Q) are constructed considering the acceptance and efficiency of the detector. Finally, we used the simultaneous fit method, as our nominal approach, to extract the $v_n$ values of the \Ds signal.

The production of prompt \Lc baryons is measured using pp and Pb-Pb collision data collected in 2017 and 2018, corresponding to integrated luminosities of 252 pb$^{-1}$ and 0.607 nb$^{-1}$, respectively. The measurements are performed in the rapidity interval $|y|<1$ and cover the \pt ranges of 3–30 GeV/c for pp and 6–40 GeV/c for Pb-Pb collisions. The raw signal yields in bins of \pt and centrality are extracted using an unbinned maximum likelihood fit to the invariant mass distributions. These raw measurements are subsequently corrected for detector acceptance and reconstruction efficiencies derived from detailed Monte Carlo simulations. The nuclear modification factor, \RAA, is then constructed from these corrected yields, defined as 

\begin{equation} 
\label{eq:raa} R_{\mathrm{AA}} = \frac{1}{\langle T_{\mathrm{AA}} \rangle} \frac{d^{2}N_{\mathrm{AA}}/dp_{\mathrm{T}}dy}{d^{2}\sigma_{\mathrm{pp}}/dp_{\mathrm{T}}dy} 
\end{equation}

where $d^{2} N_{\rm{AA}} /dp_{\rm{T}} dy$ is the differential yield of the charm hadron in Pb-Pb collisions, $d^{2} \sigma_{\rm{pp}} /dp_{\rm{T}} dy$ is the production cross section in pp collisions, and $\langle T_{AA} \rangle $ is the nuclear overlap function, which accounts for the collision geometry and is determined for each centrality class. The \Lc/\Dzero yield ratio is then determined by taking the ratio of the efficiency-corrected yields of \Lc and \Dzero mesons in the corresponding \pt and centrality bins for each collision system.

\section{Results and summary}
\label{Results}
\begin{figure}[hbt!]
\centering
\includegraphics[width=1.0\textwidth]{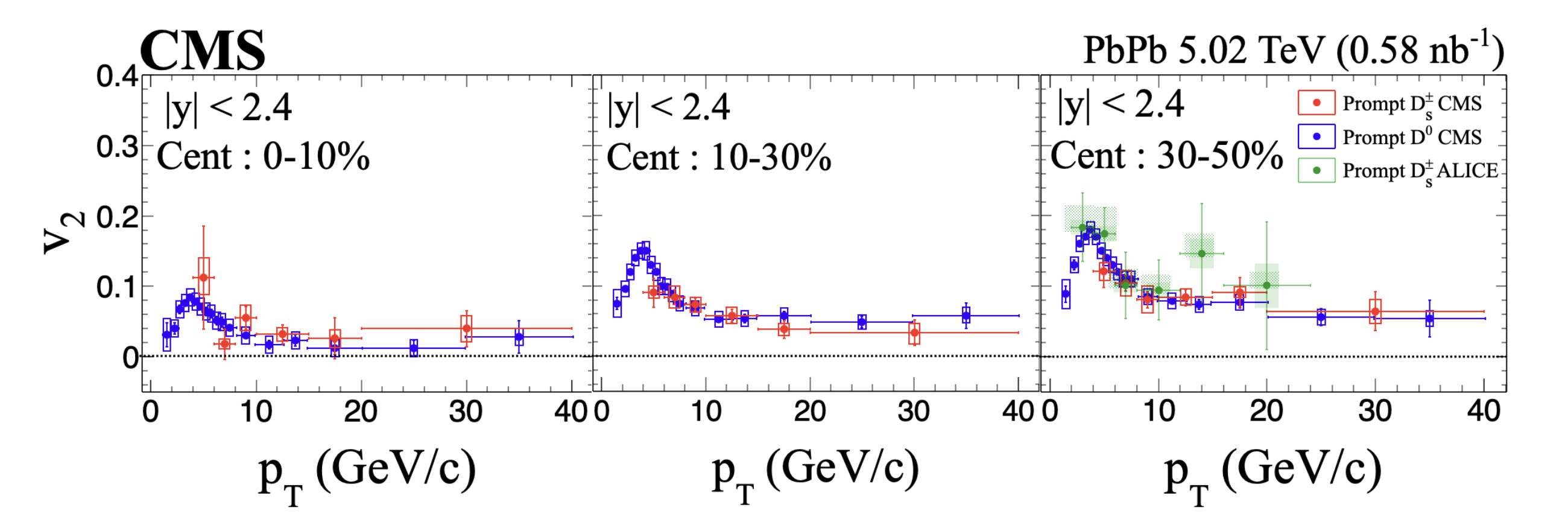}
\includegraphics[width=0.38\textwidth]{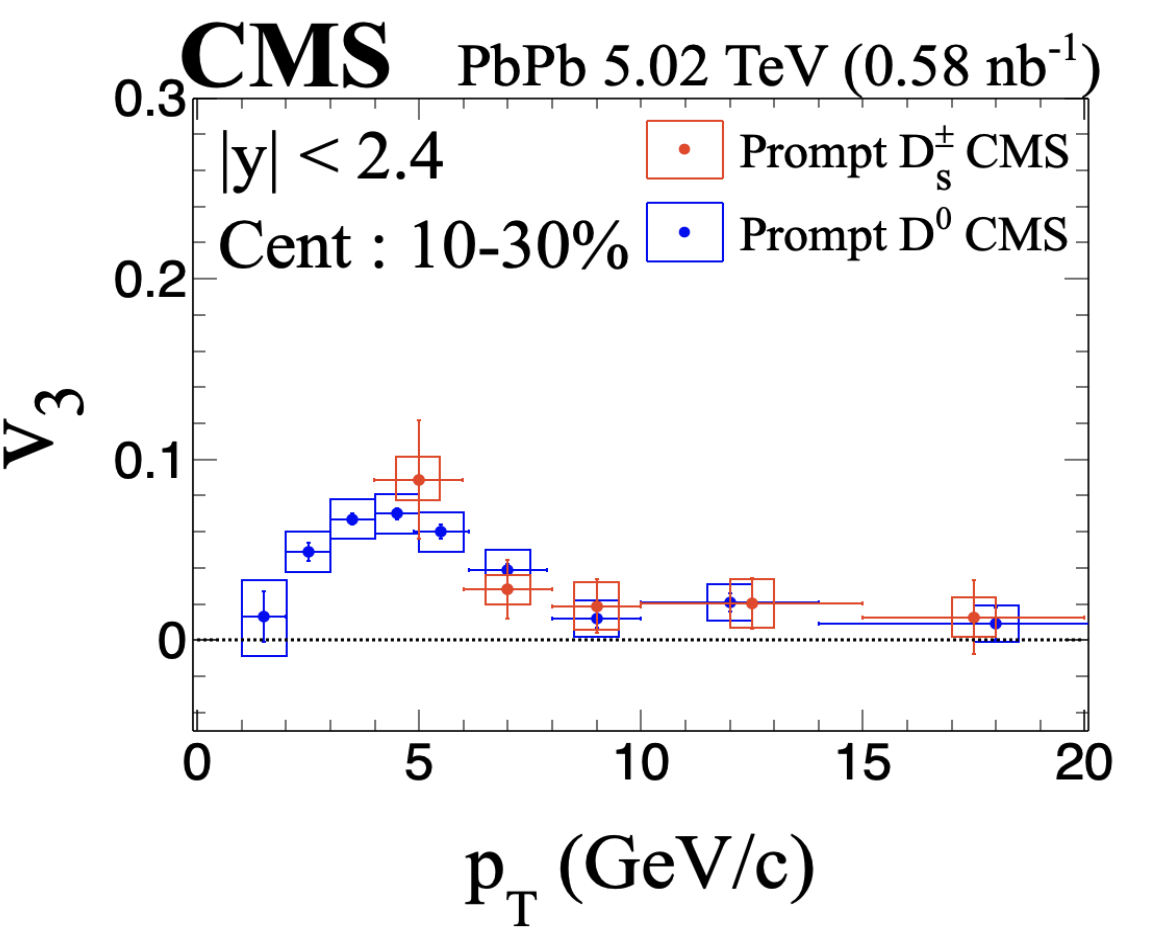}
\caption{(upper panel) The \vtwo coefficients for prompt \Ds and prompt \Dzero (from Ref.~\cite{CMS-D0-2}) mesons as a function of \pt for the 0--10\% (left), 10--30\% (middle), and 30-50\% (right) centrality classes. (lower panel)The \vthree coefficients for prompt \Ds and prompt \Dzero mesons as a function of \pt for the 10--30\% centrality class. The vertical bars and the boxes represent statistical and systematic uncertainties, respectively. The figures are from the reference~\cite{PAS-HIN-24-006}.}
\label{fig:Ds_v2}
\end{figure}

We present high-precision measurements of the elliptic (\vtwo) and triangular (\vthree) flow coefficients for prompt \Ds mesons, presented in figure~\ref{fig:Ds_v2}. The \vtwo is measured in three different centrality ranges (0-10\%, 10-30\%, and 30-50\%) spanning a \pt range of 4-40 GeV/c. However, the \vthree measurement was limited to the 10–30\% centrality bin, as large statistical fluctuations prevented a conclusive measurement in other centrality bins. These results represent a significant improvement in precision over previous studies by ALICE collaboration ~\cite{Ds_alice}. A key observation from this measurement is that the \vn of the \Ds meson is consistent with that of the \Dzero meson within uncertainties across the entire measured \pt range. This similarity suggests that the hadronization process, particularly the coalescence of a charm quark with a strange quark from the medium, does not significantly modify the azimuthal anisotropy inherited by the \Ds meson from its parent charm quark within the measured \pt range and current experimental precision ~\cite{PAS-HIN-24-006}.

\begin{figure}[hbt!]
\centering
\includegraphics[width=0.45\textwidth]{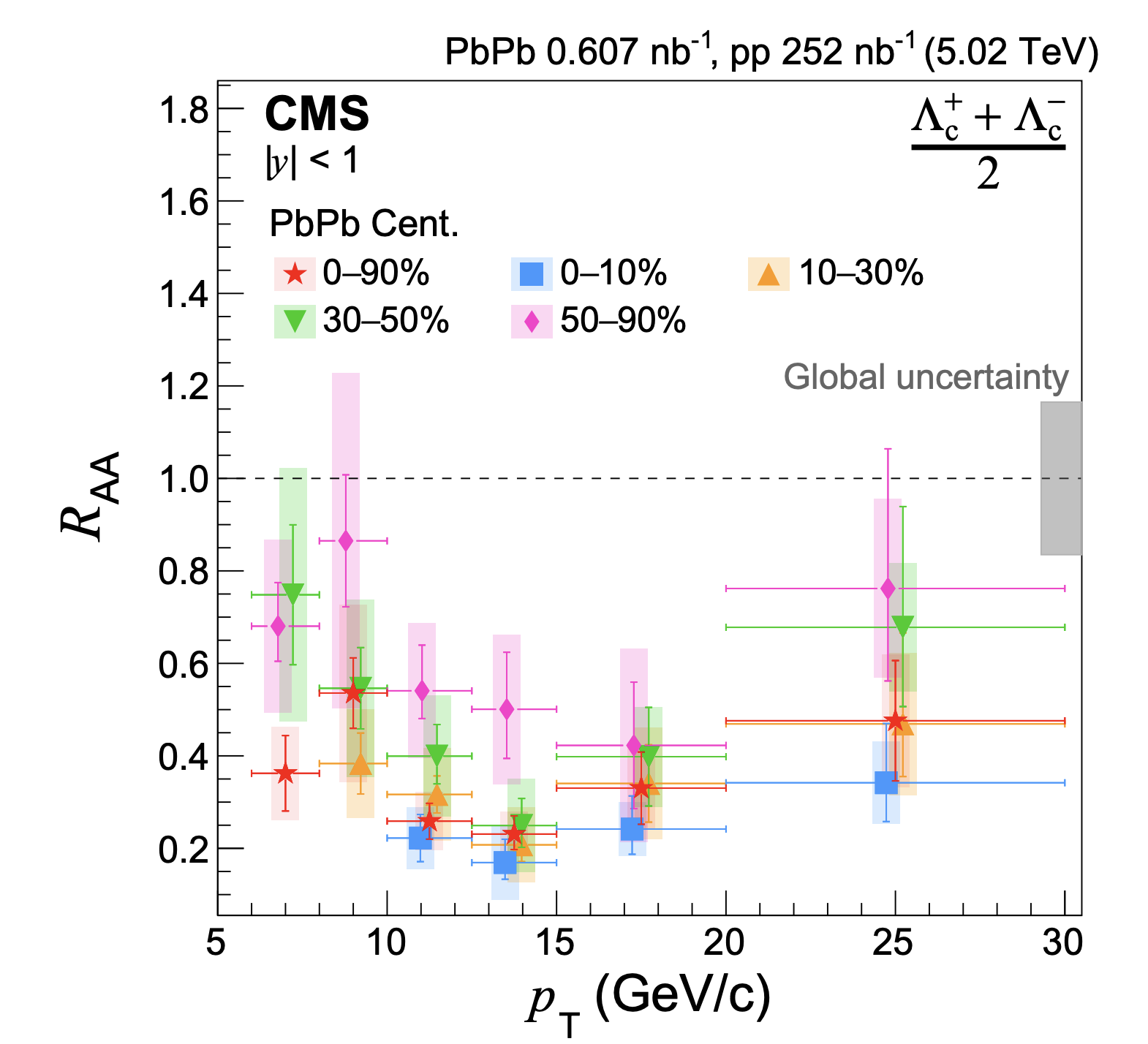}
\includegraphics[width=0.45\textwidth]{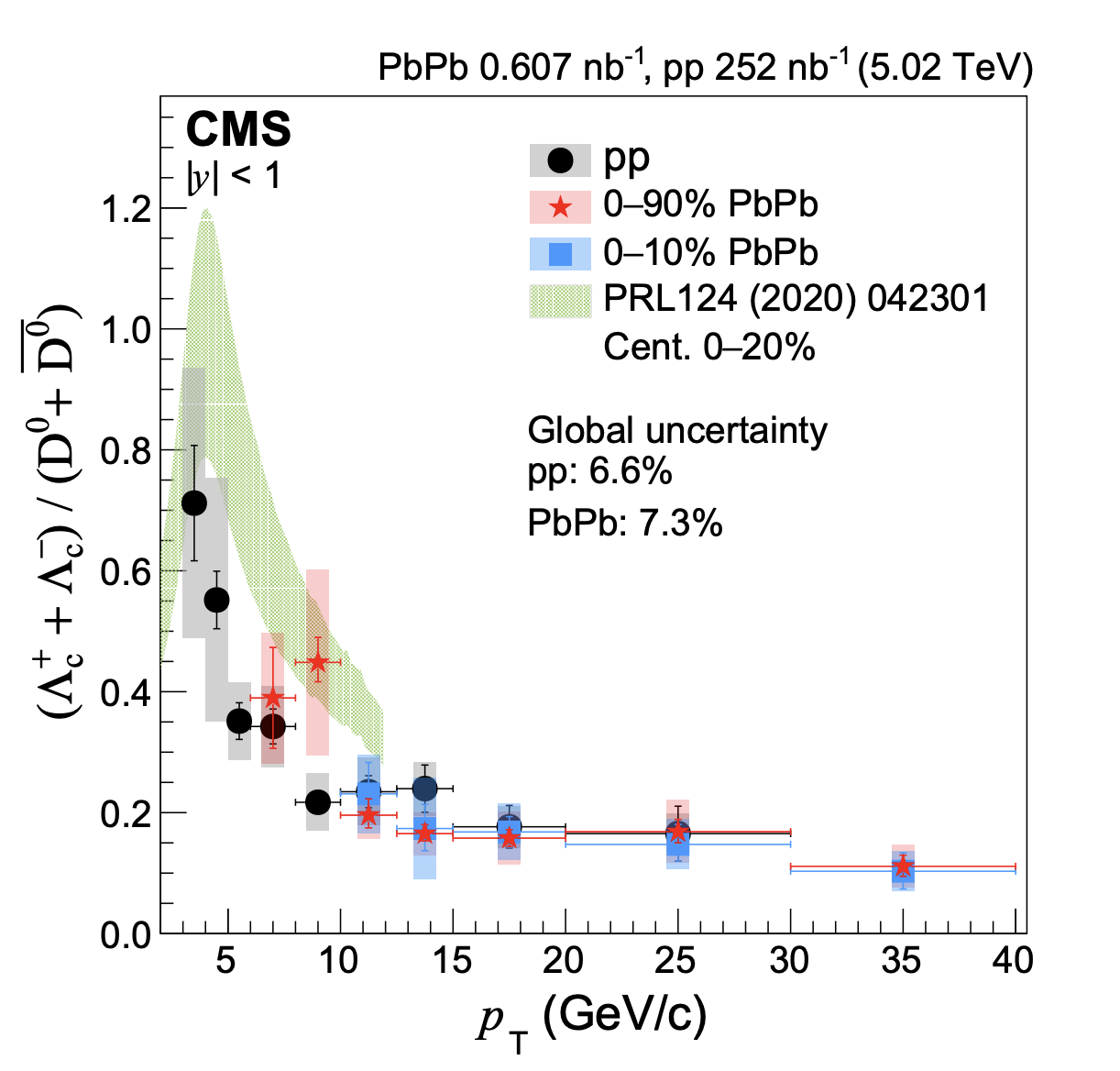}
\caption{(left) The nuclear modification factor \RAA versus \pt
for prompt \Lc production in centrality regions of 0–90\% (stars), 0–10\% (squares), 10–30\% (triangles), 30–50\% (inverted triangles) and 50–90\% (diamonds) in PbPb collisions. (right)The ratio of the production cross sections of prompt \Lc to prompt \Dzero versus \pt for 0–90\% (closed stars) and 0–10\% (closed squares) centrality classes of PbPb collisions are compared to the pp result. The vertical bars and the boxes represent statistical and systematic uncertainties, respectively. The figures are from the reference~\cite{Lc}.}
\label{fig:Lc}
\end{figure}

The \RAA measurement of \Lc, shown in the left side of figure \ref{fig:Lc}, we observe that the prompt \Lc production is found to be significantly suppressed in Pb-Pb collisions compared to pp collisions, consistent with partonic energy loss affecting charm quarks in the QGP. The suppression is larger in more central collisions and varies with the \Lc baryon \pt. The \Lc \RAA values reach their minimum near \pt $\simeq 14$ GeV/c, following the trend observed in other heavy flavor hadron measurements.

Further insight into charm hadronization is provided by the \Lc/\Dzero yield ratio across pp, pPb, and Pb-Pb collisions, shown in the right side of figure \ref{fig:Lc}. A significant enhancement of this ratio is observed in Pb-Pb collisions compared to pp collisions, particularly at intermediate \pt <10 GeV/c. This enhancement is a cornerstone prediction of coalescence models, where the abundant baryons in the QGP environment increase the probability for a charm quark to form a baryon (\Lc) rather than a meson. For \pt > 10 GeV/c, the \Lc/\Dzero ratios for pp and Pb-Pb collisions are consistent with each other, suggesting that coalescence does not play a dominant role in \Lc baryon production in this higher-\pt region. Predictions from a theoretical model (presented as a light green band) that incorporates a four-momentum conserving recombination mechanism, space-momentum correlations, and excited charm baryon decays are consistent with the data within uncertainties for the measured centrality 0-10\% and \pt $10-12.5$ GeV/c range.

\end{document}